# Screening and impurity ionization energy in semiconductors

Yuri Kornyushin

*Maître Jean Brunschvig Research Unit, Chalet Shalva, Randogne, CH-3975*

Usually microscopic electrostatic field around charged impurity ions is neglected when the ionization energy is concerned. The ionization energy is considered to be equal to that of a lonely impurity atom. Here the energy of the electrostatic field around charged impurity ions in semiconductor is taken into account. It is shown that the energy of this field contributes to decrease in the effective ionization energy. At high enough current carriers concentration the effective ionization energy becomes zero.

## 1. Introduction

Regarding impurity ionization problem it is assumed tacitly that the charge of the delocalized electrons/holes is distributed uniformly throughout a sample [1]. The impurity ionization energy at that is regarded to be equal to that of a lonely impurity atom. It is well known that a uniform distribution of current carriers in a sample is not equilibrium in the presence of the ions [1]. The charge redistributes itself, causing screening of the field of the ions [1]. This relaxation of the charge distribution leads to the decrease in the energy of a system. As a result the effective impurity ionization energy decreases also.

When $N$ randomly distributed ions and uniformly distributed charge of the current carriers are present, the electrostatic energy of a system consists of the energy of the ions [2,3], $e^2N/2\varepsilon r_0$ (here $e$ is elementary charge and $r_0$ is some small value, referring to the ion size), the energy of their charge spread uniformly through a sample [3], the energy of the uniformly spread charge of the current carriers, and the energy of the interaction of the two uniform charges. The energy of the uniformly spread ions charge, the energy of the uniformly spread charge of the current carriers and the energy of their interaction, all three of them, annihilate together because the uniformly distributed negative and positive charges compensate each other. What's left is the energy of the bare ions, $e^2N/2\varepsilon r_0$.

When the current carriers are in equilibrium, screening the ions, the electrostatic energy of a sample decreases. This means that the effective ionization energy of the impurity atoms decreases also.

## 2. Model

Let us consider a charged impurity ion in semiconductor with elementary charge $e$. This charge is positive for a donor impurity ion and negative for an acceptor one. Usually microscopic electrostatic field around charged impurity ions is neglected when the ionization energy is concerned. The ionization energy is considered to be equal to that of a lonely impurity atom [1]. Recently it was shown that the microscopic electrostatic field contributes very essentially to thermodynamic properties [2]. The matter is that when a lonely atom is ionized, the electron becomes a plane wave in an infinite or very large space. This plane wave does not posses electric field and electrostatic energy. Sometimes the analogy to the hydrogen atom is used [1]. Before the ionization of a lonely impurity atom we have electric fields of the charges of the ion and the localized electron. These fields contribute to the formation of the ground state energy of an impurity atom. The field of the electron charge does not act on electron itself, only Coulombic electric field of the ion acts. But the energy of the electrostatic field of the charge of electron contributes to formation of the energy of the ground state. After the ionization of a lonely impurity atom we have a Coulombic electric field around the ion. Its energy contributes to the formation of a quantum ground state and its energy also as was mentioned. When we have many impurity atoms ionized, we have many delocalized electrons or/and holes in the conductivity and/or valent band(s) of a semiconductor. The charge of these electrons redistributes itself, causing screening of Coulombic field of the point charge of an impurity ion and decrease in the electrostatic energy of a system.

We consider here the ions as point charges, and the delocalized electrons like a negatively charged gas.

Let us consider high temperatures, when the delocalized electrons are classical [1]. Classical delocalized electrons or/and holes screen long-range electrostatic field of point charges. The electrostatic field around a separate positive/negative ion submerged into the gas of delocalized electrons and/or holes is as follows [1]:

$$\varphi = (e/\varepsilon r)\exp{-gr}, \qquad (1)$$

where $\varepsilon$ is dielectric constant of a semiconductor, $r$ is the distance from the center of the ion and $1/g$ is Debye-Huckel/Thomas-Fermi screening radius (for classical/degenerated electrons).

The electrostatic energy of this field is smaller than that of a bare ion. So the electrostatic energy of a system is also smaller as a result of the screening. The electrostatic energy of a separate ion with electric field [Eq. (1)] is the integral over the volume of a sample of its gradient in square, divided by $8\pi$. The lower limit of the integral on $r$ should be taken as $r_0$, a very small value as was mentioned above. Otherwise the integral diverges. Calculation yields the following expression for the electrostatic energy of a separate ion [2]:

$$W = 0.5z^2e^2(r_0^{-1} + 0.5g)\exp{-2gr_0}. \qquad (2)$$

Taking into account the extreme smallness of $r_0$ and that the value of a volume is usually very large (comparative to the impurity ion volume), we have the following expression for the electrostatic energy of a separate ion [2]:

$$W = (e^2/2\varepsilon r_0) - 0.75e^2g/\varepsilon. \qquad (3)$$

Here $e^2/2\varepsilon r_0$ is the electrostatic energy of the bare ion under consideration. When $g$ is small this term is the only one in the right-hand part of Eq. (3).

The decrease in the electrostatic energy of a separate ion due to the screening is $-0.75e^2g/\varepsilon$. Electrostatic energy of $N$ randomly distributed ions is just $NW$ [3].

The effective ionization energy of the ion is correspondingly as follows:

$$W_e = W_0 - 0.75e^2g/\varepsilon, \qquad (4)$$

where $W_0$ is the initial ionization energy.

It should be mentioned here that when the screening radius is of the order of magnitude or smaller than the size of a bound state, the initial ionization energy in Eq. (4), $W_0$, is changed essentially. Here we restrict our consideration by a model, neglecting this change. We shall discuss it a bit later.

Reverse Debye-Huckel radius, $g$, increases with the increase in current carriers concentration. At some $g = g_c$ the effective ionization energy is zero. The critical value of $g$ in the accepted model is as follows:

$$g_c = 4\varepsilon W_0/3e^2. \tag{5}$$

For $\varepsilon = 11.7$ (the value for silicon) and $W_0 = 0.05$ eV Eq. (5) yields $g_c = 5.417 \times 10^6$ 1/cm (screening radius is $1.846 \times 10^{-7}$ cm.

Now let us calculate critical classical electron concentration for a donor-type semiconductor. It is well known that for this case [1]

$$g^2 = 4\pi e^2 n/\varepsilon kT, \tag{6}$$

where $n$ is current carriers concentration, $k$ is the Boltzmann constant, and $T$ is temperature. Eqs. (4), and (6) yield:

$$W_e = W_0 - (3e^3/\varepsilon)(\pi n/4\varepsilon kT)^{1/2}. \tag{7}$$

From this follows that $W = 0$ when

$$n_c = 4\varepsilon^3 kT W_0^2/9\pi e^6. \tag{8}$$

For $\varepsilon = 11.7$, $T = 300$ K and $W_0 = 0.05$ eV Eq. (8) yields $n_c = 1.964 \times 10^{19}$ 1/cm$^3$. This is rather a high concentration of delocalized electrons. It requires the same value of a concentration of donor impurity atoms (in this case average distance between the impurity atoms is $3.71 \times 10^{-7}$ cm). Atomic volume of a crystalline silicon is $v_a = 2 \times 10^{-23}$ cm$^3$. So the dimensionless critical concentration $n_c v_a$ is $3.857 \times 10^{-4}$. It is rather a large concentration, but it is much smaller than unity.

Calculations show that in the hydrogen model of the impurity atom the current carriers concentration at which the Bohr radius is equal to the Debye-Huckel radius is smaller 2.25 times than $n_c$. This justifies neglecting changes in $W_0$, while calculating $n_c$. At $n_c$ concentration the Debye-Huckel radius is still large enough (1.5 times larger than the Bohr radius) not to influence essentially the value of $W_0$.

Critical concentration decreases with decrease in temperature and sharply decreases with decrease in dielectric constant value [see Eq. (8)].

For effective mass of 10 times larger than electron mass and $n = n_c = 1.964 \times 10^{19}$ 1/cm$^3$ the Fermi energy is $2.657 \times 10^{-3}$ eV. This value is about 10 times smaller than $kT$, at $T = 300$ K, which justifies considering delocalized electrons as classical.

Effective ionization energy $W_0$ is zero at current carriers concentration equal to the critical one. This means that at $n \geq n_c$ a semiconductor regarded is no more a semiconductor, but it is a metal (or semimetal).

Eq. (8) could be rewritten in the following form:

$$T_c = 9\pi e^6 n/4k\varepsilon^3 W_0^2. \tag{9}$$

Here $T_c$ is critical temperature.

When $T \leq T_c$ the semiconductor regarded turns to be a metal/semimetal.

For degenerated delocalized electron $kT$ in Eqs. (6), (7) and (8) should be replaced by $2E_F/3$ as is well known [1]. In this case one has:

$$n_c = 0.02926\pi(\varepsilon^9 W_0^6/e^6)(\hbar^2/me^4)^3. \tag{10}$$

For effective mass of 10 times larger than free electron mass $n_c = 9.81 \times 10^{13}$ 1/cm$^3$ and the Fermi temperature is about $10^{-3}$ K. For effective mass equal to free electron mass $n_c = 9.81 \times 10^{16}$ 1/cm$^3$ and the Fermi temperature is about 1 K.

### 3. Conclusions

Only Physics of the regarded problem was analyzed here. Calculations performed are strictly valid when the current carriers concentration is essentially smaller than the critical one. To really calculate the critical concentration one needs to perform much more complicated calculations. But the model accepted is rather well founded, because the critical Debye-Huckel radius is 1.5 times larger than the radius of the quantum localized ground state of a current carrier at $n = n_c$.

The effects regarded are pronounced better for semiconductors with heavy current carriers, as corresponding Fermi temperature is smaller and the classical range is wider.